\documentclass{elsart}
\usepackage{epsfig}
\usepackage{amssymb}
\def\bea {\begin{eqnarray}}
\def\eea {\end{eqnarray}}

\def\be {\begin{equation}}
\def\ee {\end{equation}}
\begin{document} 
\begin{frontmatter} 
\title{Kaon and Lambda productions in relativistic heavy ion collisions}
\author{Jajati K. Nayak, Sarmistha Banik and Jan-e Alam}
\medskip
\address{Variable Energy Cyclotron Centre, 1/AF, Bidhan Nagar , 
Kolkata - 700064,INDIA}

\begin{abstract}
A microscopic approach has been employed to study the kaon and 
$\Lambda$ productions in heavy ion collisions. The productions
 of $K^+$ and $\Lambda$ have been studied  within the framework of Boltzmann 
transport equation for various beam energies. We find a non-monotonic horn 
like structure  for $K^+/\pi^+$ and $\Lambda/\pi$ when plotted against 
centre of mass energies ($\sqrt s_{\mathrm NN}$) with the assumption of 
initial partonic phase for $\sqrt s_{\mathrm NN}$ beyond a certain 
threshold. However, the ratio $K^+/\pi^+$ shows a monotonic nature 
when a  hadronic initial state is considered for all $\sqrt s_{\mathrm NN}$.
Experimental values of $K^-/\pi^-$ for different $\sqrt{s_{\mathrm NN}}$
are also reproduced within the ambit of the same formalism.
\end{abstract}

\end{frontmatter} 

Various signals have been proposed for the detection of quark gluon plasma
(QGP) expected to be formed in the relativistic heavy ion collisions. The
pros and cons of these signals are matter of intense debate. The strangeness
productions and its ratio to equilibrium entropy is one such widely discussed
signal. In this work we study the strangeness particularly the kaon and 
$\Lambda$ 
productions and estimate the ratio, $R^+ (\equiv K^+/\pi^+)$, 
$R^-(\equiv K^-/\pi^-)$ and $R^{\Lambda}(\equiv \Lambda/\pi)$
for various collision energies. The ratios are 
measured experimentally ~\cite{na49alt} as a function of centre of mass energy 
($\sqrt s_{NN}$) and observed that the  $R^+$ and $R^{\Lambda}$ increase with 
$\sqrt s_{NN}$ and then decrease beyond a certain value of $\sqrt s_{NN}$ 
giving rise to a horn like structure, whereas the ratio, $R^-$ increases 
monotonically, faster at lower $\sqrt s_{NN}$ and tend to saturate at 
higher $\sqrt s_{NN}$. 
Various models ~\cite{theory} have been proposed in the literature
to explain the data. 
In this work the kaon and $\Lambda$ productions have been calculated for 
various $\sqrt s_{NN}$, ranging from 3.32 to 200 GeV and
 examine whether the $K^+/\pi^+$ experimental 
data can differentiate the following scenarios of the system that is 
assumed to be formed after collision. The system is formed; (I) in the 
hadronic phase for all 
$\sqrt s_{\mathrm {NN}}$ or (II) in the partonic phase beyond a certain 
threshold in $\sqrt s_{\mathrm {NN}}$. Other possibilities like formation of 
the system with strangeness in complete thermal equilibrium and the evolution
in space time (III) without and (IV) with secondary productions
of quarks and hadrons have been considered. (V) Results for an extreme 
case of zero strangeness in the initial state is also discussed.

We consider the processes of gluon-gluon fusion and light quarks annihilation 
for the strangeness production ($s$,$\bar s$) in the partonic phase. For the 
strangeness production ($K^+$, $K^-$ and $\Lambda$) in hadronic phase an 
exhaustive set of reactions involving thermal baryons and mesons have been 
considered ~\cite{randrup,brown2}. 
The possibility of formation of a fully equilibrated system  in high
energy nuclear collisions is still a fiercely debated issue because of
the finite size and life time of the system. In scenario-I \& II we assume
that the strange quarks or the strange hadrons (depending on the value of
$\sqrt s_{\mathrm {NN}}$) are produced out of chemical equilibrium and the 
non-strange quarks and hadrons are in complete thermal equilibrium. Therefore, 
the evolution of the strange sector of the system is governed by the 
interactions between the equilibrium and non-equilibrium degrees of freedom. 
In such cases the time evolution of the densities of the strange quarks and 
hadrons can be evaluated by the momentum integrated Boltzmann equation which 
reads, 
\begin{figure}
\begin{center}
\includegraphics[scale=0.3]{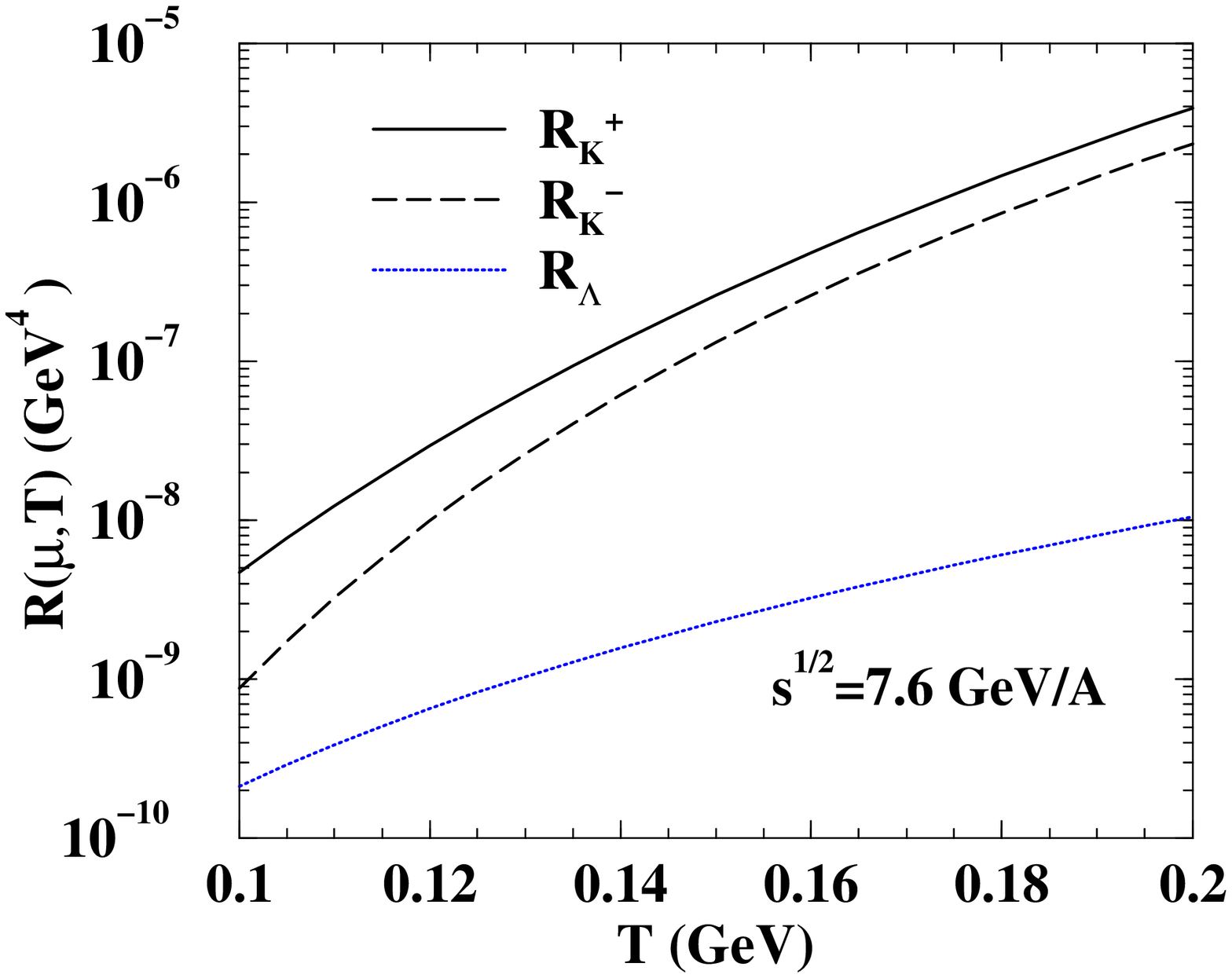}
\includegraphics[scale=0.3]{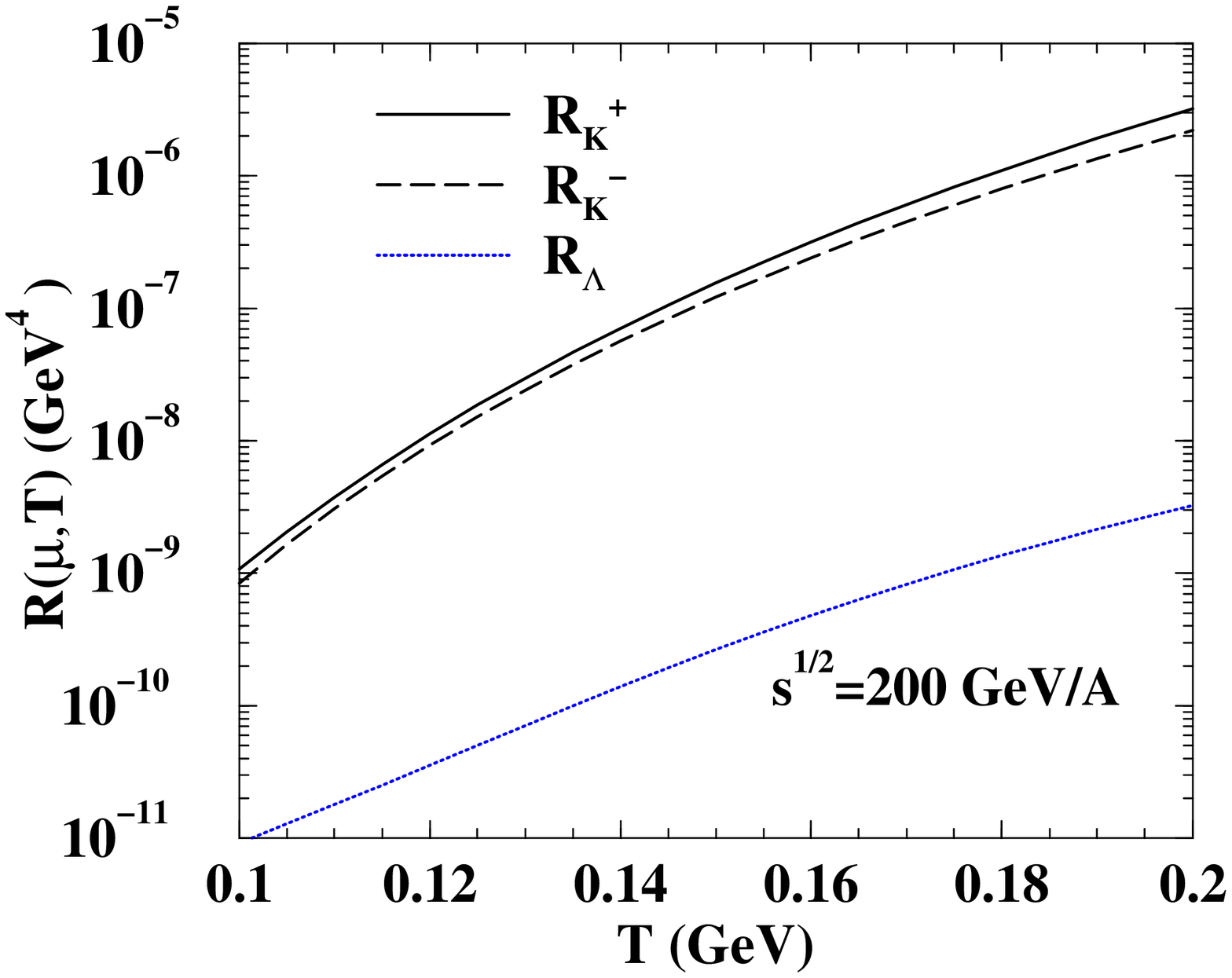}
\caption{Rate of production of $K^+$, $K^-$ and $\Lambda$ for 
$\sqrt{s_{\mathrm NN}}$=7.6 (left) and 200 GeV (right).}
\label{fig1}
\end{center}
\end{figure}
\begin{eqnarray}
\frac{dn_{i,j}}{d\tau} &=& R_{i,j}(\mu_B, T)
\left[1 - \frac{n_i n_j}{n_{i}^{eq}{n_{j}^{eq}}}\right] - \frac{n_{i,j}}
{\tau}. 
\label{eqn_evol1}
\end{eqnarray}
where, $n_i$ ($n_j$) and ${n_i}^{eq}$ (${n_j}^{eq}$) are the non-equilibrium
and equilibrium densities of $i$ ($j$) type of particles respectively.
$R_i$ is the rate of production of particle $i$ at temperature $T$ and 
chemical potential $\mu_B$, $\tau$ is the proper time. First term on the 
right hand side of Eq. \ref {eqn_evol1} is the net production term and the 
second term represents  the dilution of the system due to expansion. 
The indices $i$ and $j$ in Eq.\ref{eqn_evol1} 
are replaced by $s, \bar s$ quark in the QGP phase  
and by $K^+, K^-$ in the hadron phase 
respectively. For $\sqrt s_{NN} \le $ 7.6 GeV an
initial hadronic states and for 
$\sqrt s_{NN} > $ 7.6 GeV an initial partonic phase is assumed  
(scenarios-II,III,IV and V). In these scenarios the hadrons are 
assumed to be formed at a transition temperature, $T_c=190$ MeV through a 
first order phase transition. In case of a first order transition the 
evolution of kaons in the mixed phases is given by;  
\begin{eqnarray}
\frac{dn_{K^{\pm}}}{d\tau}& = &R_{K^{\pm}}(\mu_B,T_c)
[ 1 - \frac{n_{K^{\pm}}n_{K^{\mp}}}{n_{K^{\pm}}^{eq}n_{K^{\mp}}^{eq}}] - 
\frac{n_{K^{\pm}}}{\tau}+ \frac {1}{f_H} \frac {d f_H}{d \tau} 
\left(\delta n_{\bar{s} (s)}-n_{K^{\pm}}\right) 
\label{eqmix}
\end{eqnarray}
The last term in the Eq.~\ref{eqmix} is the hadronisation term and 
$f_H(\tau)=1-f_Q(\tau)$ represents the fraction of hadrons in the mixed
phase at time $\tau$. $f_Q(\tau)=(1/(r-1))(r \tau_H/\tau-1)$, is the 
fraction of the QGP phase. $\tau_Q$ ($\tau_H$) is the time at 
which the QGP (mixed) phase ends, $r$ is the ratio of the statistical 
degeneracies in QGP to hadronic phase. Similar equations exist for the 
evolution of $s$ and $\bar{s}$ quarks in the mixed phase~\cite{kapusta}.
The $\delta$ in Eq.~\ref{eqmix}is a parameter which indicates 
the fraction of $\bar{s}(s)$  quarks hadronizing to $K^+(K^-)$. In the same 
approach the $\Lambda$ productions have also been calculated for all the 
$\sqrt s_{NN}$. The rates of production for $K^+, K^-$ and $\Lambda$ for 
$\sqrt s_{NN}$=7.6 and 200 GeV are displayed in Fig.~\ref{fig1} (for the 
strange quark productions we refer to~\cite{jkn}). The rate of productions of 
$K^+$ and $K^-$ are similar at large $\sqrt{s_{\mathrm NN}}$ (low $\mu_B$) 
although at smaller $\sqrt{s_{\mathrm NN}}$ (high $\mu_B$) $K^-$ production 
is smaller because of the larger $K^-$-nucleon absorption cross section at 
non-zero baryon density. The $\Lambda$ yield is much smaller because of 
smaller production cross section.
\begin{figure}
\begin{center}
\includegraphics[scale=0.3]{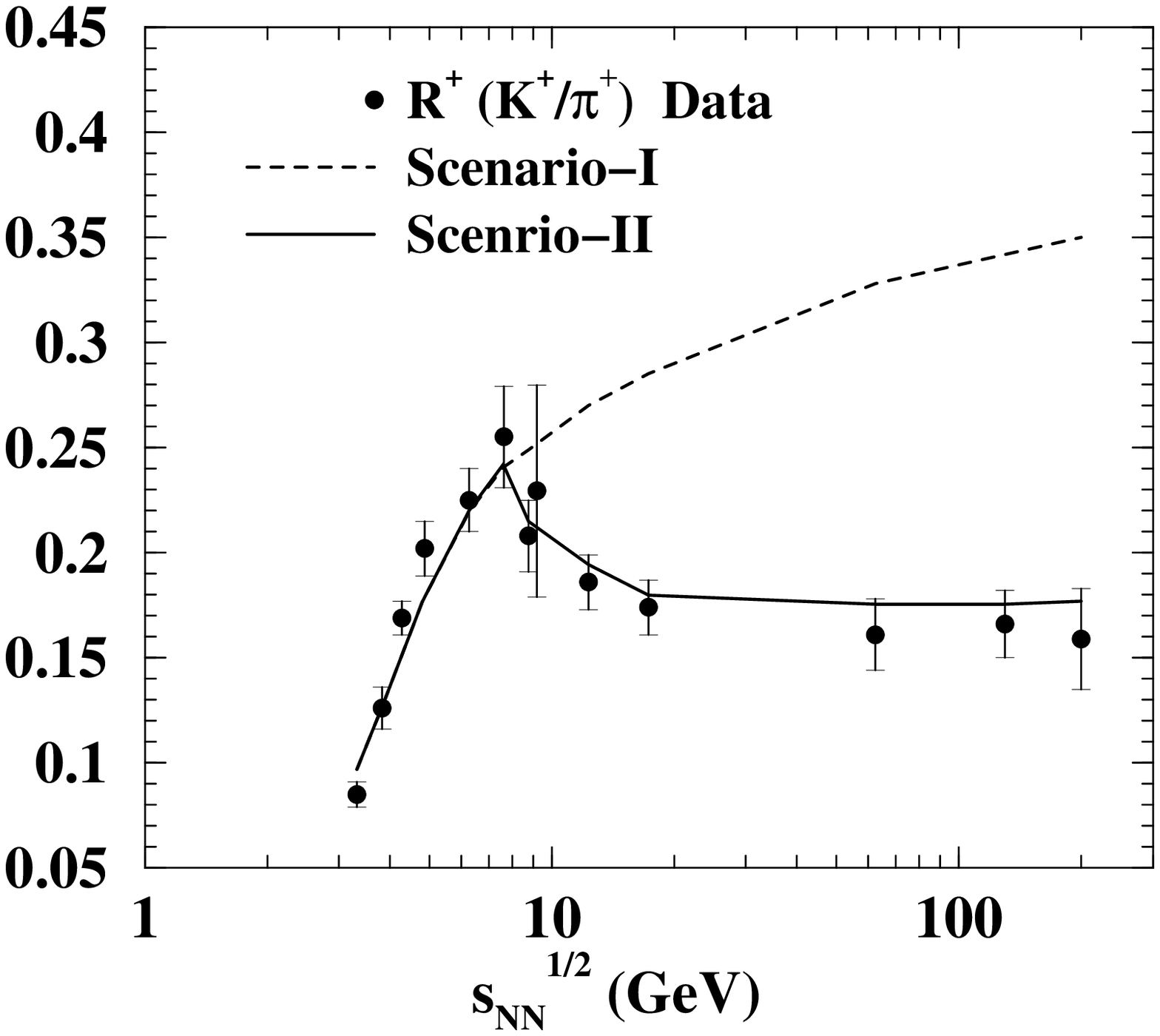}
\includegraphics[scale=0.3]{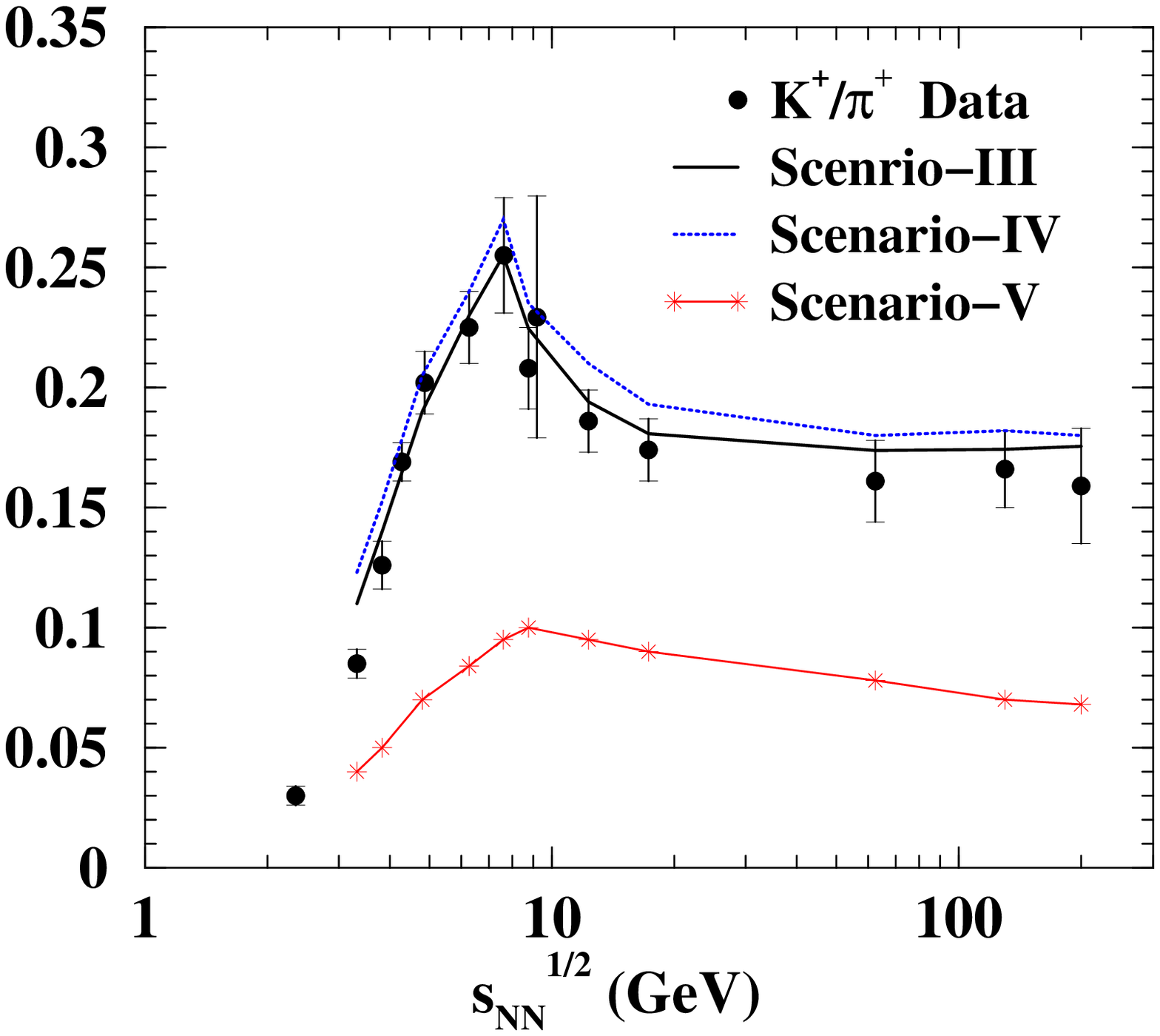}
\caption{The variation of $R^+$ with $\sqrt s_{\mathrm NN}$ for different 
scenarios (see text).}
\label{fig2}
\end{center}
\end{figure}
The baryonic chemical potential at freeze-out are taken from the 
parametrization of $\mu_B$ with $\sqrt{s_{\mathrm NN}}$ (see the 
table in~\cite{jkn} and the references there in) and 
its value at the initial state is obtained from the net
baryon number conservation equation. The variation of temperature
and net baryon density has been obtained from the solution of
boost invariant relativistic hydrodynamics~\cite{bjorken}.

Results on  $K^+$/$\pi^+$  for scenarios-I and II are displayed in 
Fig.~\ref{fig2} (left panel). In scenario-I a monotonic rise of $R^+$ for 
all $\sqrt{s_{\mathrm NN}}$ is observed. Whereas in scenario-II a 
non-monotonic horn like structure for $R^+$ is observed  which describes 
the experimental data reasonably well. Now we consider the scenarios III, IV 
and V (Fig.~\ref{fig2}, right). On one hand when the strangeness evolve from 
the initial state till the freeze-out stage in complete thermal equilibrium 
and the secondary productions of strange quarks and hadrons are off (III) the 
data is reproduced well. On the other hand the same equilibrium scenario 
overestimates  the data slightly at high $\sqrt s_{\mathrm NN}$ if the 
secondary production is switched on (IV). This indicates that the deficiency 
of strangeness below its equilibrium value as considered in (II) is 
compensated by the secondary productions. In scenario V we assume that 
vanishing initial strangeness and observed that the production of strangeness 
throughout the evolution is not sufficient to reproduce the data. 
\begin{figure}
\begin{center}
\includegraphics[scale=0.3]{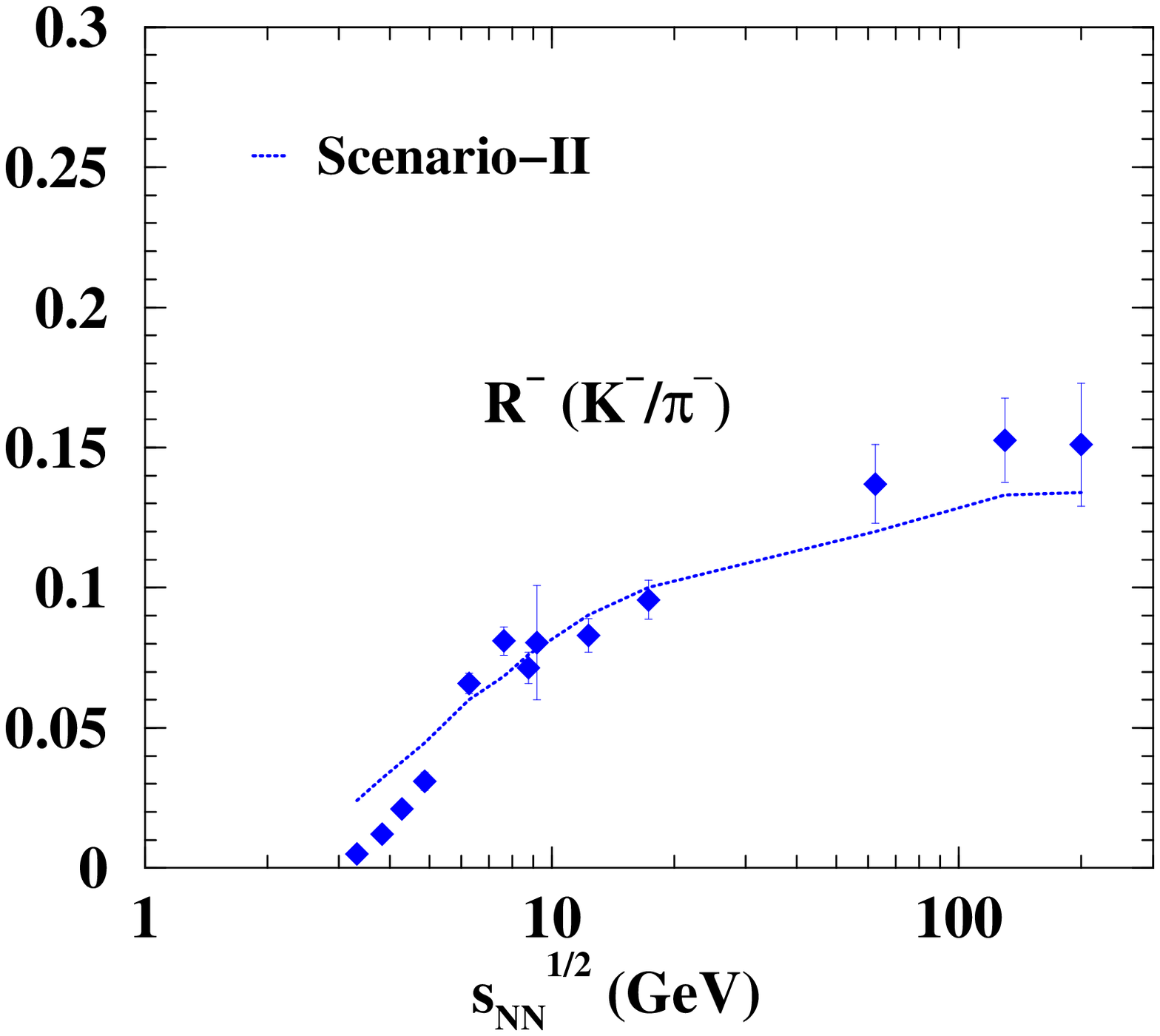}
\includegraphics[scale=0.3]{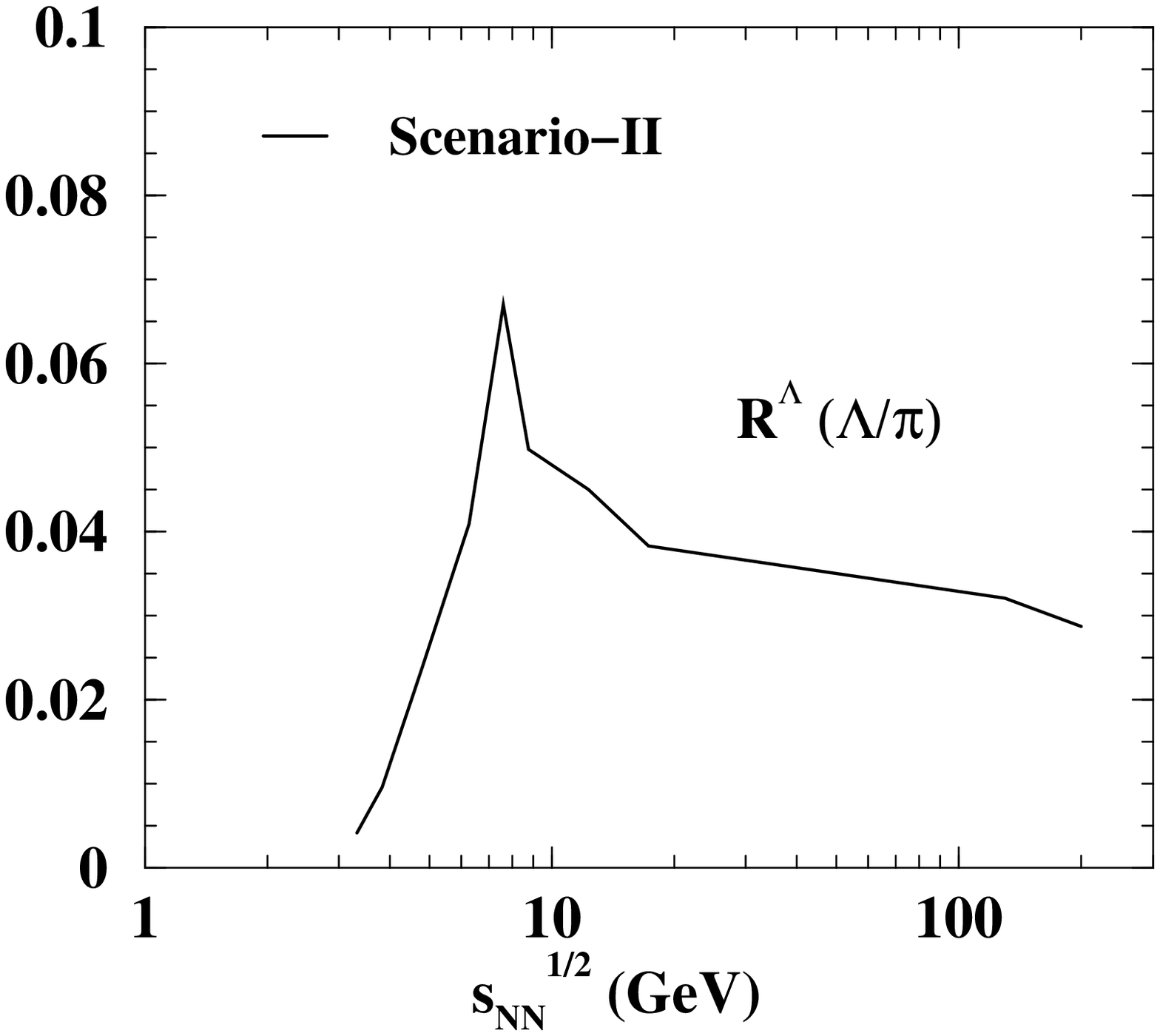}
\caption{Variation of $R^-$ (left)  and $R^{\Lambda}$ (right) for scenario-II.  
}
\label{fig3}
\end{center}
\end{figure}
In Fig.~\ref{fig3} the $R^-$ has been displayed as a function of
$\sqrt s_{\mathrm NN}$ in the left panel of the curve for scenario-II. In the 
right panel the $R^{\Lambda}$ is shown for the same 
scenario of initial condition. Theoretically a horn like structure 
is obtained for $\Lambda$ which resembles the experimental data~\cite{blume}.  

In summary, we have studied the kaon and $\Lambda$ production in heavy ion
collisions for various collision energies. The analysis of the experimental
data indicates the formation of a partonic state for $\sqrt{s_{\mathrm NN}}$
above 7.6 GeV.
 
{\bf Acknowledgment:} 
S. Banik and J.A  are supported by DAE-BRNS project Sanction No.
2005/21/5-BRNS/2455. Authors thank C. Blume and B. Mohanty for providing 
useful documents.

\end{document}